\def\CMP{\sevenrm Commun.\ Math.\ Phys.}

\def\LMP{\sevenrm Lett.\ Math.\ Phys.}

%
%
\def\today{\number\day .\space\ifcase\month\or
January\or February\or March\or April\or May\or June\or
July\or August\or September\or October\or November\or December\fi, \number \year}
%
%
\newcount \theoremnumber
\def\cleartheoremnumber{\theoremnumber = 0 \relax}

\def\Prop #1 {
             \advance \theoremnumber by 1
             \vskip .6cm 
             \goodbreak 
             \noindent
             {\bf Proposition III.{\the\theoremnumber}.}
             {\sl #1}  \goodbreak \vskip.8cm}

\def\Conj#1 {
             \advance \theoremnumber by 1
             \vskip .6cm  
             \goodbreak 
             \noindent
             {\bf Conjecture {\the\headlinenumber}.{\the\theoremnumber}.}
             {\sl #1}  \goodbreak \vskip.8cm} 

\def\Th#1 {
             \advance \theoremnumber by 1
             \vskip .6cm  
             \goodbreak 
             \noindent
             {\bf Theorem III.{\the\theoremnumber}.}
             {\sl #1}  \goodbreak \vskip.8cm}

\def\Lm#1 {
             \advance \theoremnumber by 1
             \vskip .6cm  
             \goodbreak 
             \noindent
             {\bf Lemma III.{\the\theoremnumber}.}
             {\sl #1}  \goodbreak \vskip.8cm}

\def\Cor#1 {
             \advance \theoremnumber by 1
             \vskip .6cm  
             \goodbreak 
             \noindent
             {\bf Corollary III.{\the\theoremnumber}.}
             {\sl #1}  \goodbreak \vskip.8cm} 
%
%
\newcount \equationnumber

\newcount \refnumber

\def\[]    {\global 
            \advance \refnumber by 1
            [{\the\refnumber}]}

\def\# #1  {\global 
            \advance \equationnumber by 1
            $$ #1 \eqno ({\the\equationnumber}) $$ }

\def\% #1 { \global
            \advance \equationnumber by 1
            $$ \displaylines{ #1 \hfill \llap ({\the\equationnumber}) \cr}$$} 

\def\& #1 { \global
            \advance \equationnumber by 1
            $$ \eqalignno{ #1 & ({\the\equationnumber}) \cr}$$}
%
%
\newcount \Refnumber

\def\Ref #1 #2 #3 #4 #5 #6  {\ninerm \global
                             \advance \Refnumber by 1
                             {\ninerm #1,} 
                             {\ninesl #2,} 
                             {\ninerm #3.} 
                             {\ninebf #4,} 
                             {\ninerm #5,} 
                             {\ninerm (#6)}\nobreak} 
\def\Bookk #1 #2 #3 #4       {\ninerm \global
                             \advance \Refnumber by 1
                             {\ninerm #1,}
                             {\ninesl #2,} 
                             {\ninerm #3,} 
                             {(#4)}}
\def\Book{\cr
{\the\Refnumber} &
\Bookk}
\def\Reff{\cr
{\the\Refnumber} &
\Ref}
\def\REF #1 #2 #3 #4 #5 #6 #7   {{\sevenbf [#1]}  & \hskip -9.5cm \vtop {
                                {\sevenrm #2,} 
                                {\sevensl #3,} 
                                {\sevenrm #4} 
                                {\sevenbf #5,} 
                                {\sevenrm #6} 
                                {\sevenrm (#7)}}\cr}
\def\BOOK #1 #2 #3 #4  #5   {{\sevenbf [#1]}  & \hskip -9.5cm \vtop {
                             {\sevenrm #2,}
                             {\sevensl #3,} 
                             {\sevenrm #4,} 
                             {\sevenrm #5.}}\cr}
\def\HEP #1 #2 #3 #4     {{\sevenbf [#1]}  & \hskip -9.5cm \vtop {
                             {\sevenrm #2,}
                             {\sevensl #3,} 
                             {\sevenrm #4.}}\cr}
%
%
\def\bull{$\sqcup \kern -0.645em \sqcap$}
%
%
\def\DefII#1{  \advance \theoremnumber by 1
             \vskip .6cm  
             \goodbreak 
             \noindent
             {\bf Definition II.{\the\theoremnumber}.}
             {\sl #1}  \goodbreak \vskip.4cm}
\def\DefIII#1{  \advance \theoremnumber by 1
             \vskip .6cm  
             \goodbreak 
             \noindent
             {\bf Definition III.{\the\theoremnumber}.}
             {\sl #1}  \goodbreak \vskip.4cm}
\def\Rem#1{\vskip .4cm \goodbreak \noindent
                                     {\it Remark.} #1 \goodbreak \vskip.5cm }

\def\Pr#1{\goodbreak \noindent {\it Proof.} #1 \hfill \bull  \goodbreak \vskip.5cm}

%
%
\def\*{\vskip 1.0cm}      

%
%
\newcount \ssubheadlinenumber

\def\SSHL #1 {\goodbreak
            \cleartheoremnumber
            \vskip 1cm
            \advance \ssubheadlinenumber by 1
{\rm \noindent {\the\headlinenumber}.{\the\subheadlinenumber}.{\the\ssubheadlinenumber}. #1}
            \nobreak \vskip.8cm \rm \noindent}
\newcount \subheadlinenumber
\def\clearsubheadlinenumber{\subheadlinenumber = 0 \relax}
\def\SHL #1 {\goodbreak
            \cleartheoremnumber
            \vskip 1cm
            \advance \subheadlinenumber by 1
            {\rm \noindent {\the\headlinenumber}.{\the\subheadlinenumber}. #1}
            \nobreak \vskip.8cm \rm \noindent}
\newcount \headlinenumber

\newcount \headlinesubnumber
\def\clearheadlinesubnumber{\headlinesubnumber = 0 \relax}
\def\Hl #1 {\goodbreak
            \cleartheoremnumber
            \clearheadlinesubnumber
            \clearsubheadlinenumber
            \advance \headlinenumber by 1
            {\bf \noindent #1}
            \nobreak \vskip.4cm \rm \noindent}
\font\twentyrm=cmr17
\font\fourteenrm=cmr10 at 14pt
\font\sevensl=cmsl10 at 7pt
\font\sevenit=cmti7 

\font\css=cmss10
\font\Rosch=cmr10 at 9.85pt
\font\Cosch=cmss12 at 9.5pt
\font\rosch=cmr10 at 7.00pt
\font\cosch=cmss12 at 7.00pt
\font\nosch=cmr10 at 7.00pt
%
%
%
%
%
%
%
%
%
%
%
%
%
%
%
%
\def\Z                 {\hbox{{\css Z}  \kern -1.1em {\css Z} \kern -.2em }}
\def\R                 {\hbox{\raise .03ex \hbox{\Rosch I} \kern -.55em {\rm R}}}
\def\N                 {\hbox{\rm I \kern -.55em N}}
\def\C                 {\hbox{\kern .20em \raise .03ex \hbox{\Cosch I} \kern -.80em {\rm C}}}

\def\r                 {\hbox{\raise .03ex \hbox{\rosch I} \kern -.45em \hbox{\rosch R}}}
\def\n                 {\hbox{\hbox{\rosch I} \kern -.45em \hbox{\nosch N}}}
\def\c                 {\hbox{\raise .03ex \hbox{\cosch I} \kern -.70em \hbox{\rosch C}}}

\def\z                 {\hbox{\kern 0.2em {\cal z}  \kern -0.6em {\cal z} \kern -0.3em  }}
\def\1                 {\hbox{\rm \thinspace \thinspace \thinspace \thinspace
                                  \kern -.50em  l \kern -.85em 1}}
\def\unit                 {\hbox{\sevenrm \thinspace \thinspace \thinspace \thinspace
                                  \kern -.50em  l \kern -.85em 1}}
%
%
%
%
%
%
%
%
%
%
%
%

\def\A                 {{\cal A}}

\def\B                 {{\cal B}} 

\def\H                 {{\cal H}} 
 
\def\O                 {{\cal O}}

%

%
%
%
%
%
%
%


\def\versuch #1 #2 {
\vskip -.1 cm
\global \advance \equationnumber by 1
            $$\displaylines{ \rlap{ #1 } \hfill #2  \hfill \llap{({\the\equationnumber})} } $$ 
\vskip  .1cm
\noindent}

\nopagenumbers
\def\Draft  {\hbox{Preprint \today}}
\def\firstheadline{\hss \hfill  \Draft  \hss} 
\headline={
\ifnum\pageno=1 \firstheadline
\else 
\ifodd\pageno \rightheadline 
\else \leftheadline \fi \fi}
\def\rightheadline{\sevenrm THE REEH-SCHLIEDER PROPERTY FOR THERMAL FIELD THEORIES
\hfill \folio } 
\def\leftheadline{\sevenrm \folio \hfill CHRISTIAN D.\ J\"AKEL}
\voffset=2\baselineskip
\magnification=1200
%
%
%
%
%

\vskip 1cm

\noindent
{\twentyrm The Reeh-Schlieder Property for Thermal Field Theories}

\vskip 1cm
\noindent
{\sevenrm CHRISTIAN D.\ J\"AKEL}

\noindent
{\sevenit  Dipartimento di Matematica,
Via della Ricerca Scientifica, Universit\`a di Roma ``Tor Vergata'', 
I-00133~Roma, e-mail: christian.jaekel@uibk.ac.at}

\vskip .5cm     
\noindent {\sevenbf Abstract}. {\sevenrm We show that 
the Reeh-Schlieder property w.r.t.\ KMS states is a direct consequence of 
locality, additivity and the relativistic KMS condition. The latter characterizes
the thermal equilibrium states of a relativistic quantum field theory.
The statement remains valid even if the given equilibrium state breaks spatial 
translation invariance.}

\vskip 1 cm

%
%
%
%
%

\Hl{I.\ Introduction}

\noindent
In a relativistic quantum theory (even of massive particles), 
the vacuum state, which is simply characterized by Poincar\'e invariance and 
the spectrum condition, has a very rich intrinsic structure. 
The full content of the theory can be described in terms of vacuum expectation values.
And there are more surprises out there. While studying a relativistic quantum theory and 
its vacuum state, one encounters a number of most peculiar properties, which require 
a drastic departure from `classical' quantum mechanics and its interpretation. 
The famous Reeh-Schlieder property may be seen as one of the origins of
these peculiarities; with astonishing consequences for the theory and its interpretation. 
It somehow exploits the correlations between the vacuum expectation values of measurements
in spacelike separated regions. According to the celebrated Cluster Theorem 
these correlations decay exponentially (as long as the theory describes only massive particles). 
Therefore the energy necessary to exploit them puts severe limits on the size of 
affortable effects. Nevertheless, the Reeh-Schlieder Theorem states that if there where no 
restrictions on the 
available energy, then one could prepare any vector state with arbitrary accuracy using only 
strictly local operations; i.e., operations 
performed in an arbitray bounded space--time region. (In the mathematical description, 
the vacuum state is specified by a vector 
$\Omega$ in the so-called vacuum Hilbert space $\H$ 
and a local operation is modelled by a linear operator acing on $\H$.)
 
The Reeh-Schlieder property for thermal equilibrium  states was first proven by Jung\-las.$^1$ 
He assumed that the thermal equilibrium  state is locally normal w.r.t.\ the vacuum and
his proof relied on a result of Borchers concerning timelike cylinders in the 
vacuum representation. Only time translations were used and therefore the (standard)
KMS condition, which characterizes equilibrium states, was sufficient. 
Here we present a self-contained derivation of the Reeh-Schlieder property,
which does not rely on results concerning the vacuum sector, but instead takes 
advantage of the  {\sl relativistic KMS condition}, recently proposed by
Bros and Buchholz.$^2$ We will introduce it in the next section. 
We would like to emphasize that we  {$\underline {\hbox{do  not}}$}  require that there 
exists a group of unitary operators which implements spacelike translations in the thermal 
representation associated  with a given equilibrium  state. In fact, our proof remains applicable
even if the thermal state breaks translation or rotation symmetry. 
It uses Glaser's Theorem and exploits the characteristic analyticity properties 
of an equilibrium state. The latter simply reflect the basic stability and passivity 
properties of an equilibrium state. 

\Rem{Recently the Cluster Theorem has been generalized to thermal states.$^3$
By simply combining the KMS condition with the locality assumption the author was able 
to show that there is a tight relation between the infrared properties of the generator 
of time translations and the decay of spatial
correlations in any extremal KMS state, in complete analogy to the 
well understood case of the vacuum state. To be more precise, since the spectrum of the
generator of the time-evolution in the thermal sector does not have a mass gap, 
a new condition proposed by Buchholz, which may be interpreted as a type of H\"older 
continuity of the spectrum 
at the discrete eigenvalue zero, has been used to show that the
correlations between two spacelike separated measurements
decay like some inverse power of their spatial distance. The correlations of free massless
bosons in two dimensions saturate these bounds.}

To conclude this introduction we line out the content of this paper. In Section 2
basic properties of thermal quantum field theories and their representations
are collected, including the relativistic KMS condition of Bros and Buchholz. 
Section 3 contains the derivation of the Reeh-Schlieder property for thermal 
equilibrium states. A brief outlook is given in the final section.

\vskip 1cm

\Hl{II.\ Thermal Quantum Field Theory}

\noindent
In the algebraic formulation$^4$ a QFT is casted into an inclusion preserving map 
\# {\O \to \A (\O)} 
which assigns to any open bounded region $\O$ in Minkowski 
space $\R^4$ a unital $C^*$-al\-gebra~$\A (\O)$. 
The Hermitian elements of the {\sl abstract} $C^*$-algebra $\A (\O)$ are interpreted as 
the observables which can be measured at times and locations in $\O$. 
The net $\O \to \A(\O)$ is isotonous, i.e., there exists a unital embedding
\# {\A(\O_1) \hookrightarrow \A(\O_2)
\qquad \hbox{if} \quad \O_1 \subset \O_2.} 
For mathematical convenience the local algebras are embedded in the $C^*$-inductive limit 
algebra
\# { \A = \overline{ \cup_{ {\cal O}  \subset \r^4} \A(\O) }^{\, C^*}  .}  
The space--time symmetry of Minkowski space manifests itself
in the existence of a representation 
\# {\alpha \colon ( \Lambda, x) \mapsto \alpha_{ \Lambda, x} \in Aut (\A), 
\qquad  (\Lambda, x) \in {\cal P}_+^\uparrow ,} 
of the (orthochronous)
Poincar\'e group ${\cal P}_+^\uparrow$. Lorentz-transformations $\Lambda$  and space--time 
translations~$x$ act geometrically:
\# {\alpha_{ \Lambda, x} \bigl( \A (\O) \bigr) 
= \A (\Lambda \O + x)  \qquad \forall (\Lambda, x) \in {\cal P}_+^\uparrow.} 
Einstein causality is implemented by locality:
observables localized in spacelike separated space--time regions commute, i.e.,
\# { \A (\O_1) \subset \A^c ( \O_2) \quad \hbox{\rm if} \quad \O_1 \subset \O_2'.}
Here $\O '$ denotes the spacelike complement of $\O$ and 
$\A^c (\O)$ denotes the set of operators in~$\A$ which commute with all operators in $\A(\O)$.

\Rem{Let $h \in L^1 (\R^4 , {\rm d}^4 x)$  such that the Fourier-transform
$\tilde h$ of $h$ has compact support. Strong continuity of the group of
automorphisms  $x \mapsto \alpha_x$ implies that the Bochner integral 
\# {a_h = \int {\rm d}^4 x \, \, h(x) \alpha_x (a) ,  \qquad a \in \A , }
exists in $\A$ and defines an entire analytic element for the translations. 
(If the map $x \mapsto \alpha_x$ fails to be strongly continuous, then
we may proceed by simply restricting the given net 
$\O \to \A(\O)$ to the subnet consisting of those
elements of $\A$ which comply with the continuity 
condition (see, e.g., Ref.\ 5, Proposition 1.18).) Recall that
$b \in \A$ is called an entire analytic element for the group of automorphisms 
$x \mapsto \alpha_x$, if there exists a function
$g \colon \C^4 \to \A$ such that
\vskip .2cm
\hskip .5cm
i) $\qquad g(x) = \alpha_x (b)$ for all $ x \in \R^4$ ; 
\vskip .2cm
\hskip .5cm
ii) $\qquad z \mapsto \omega \bigl( g(z) \bigr)$ 
is entire analytic for all positive linear funtionals $\omega$ over ${\cal A}$.
\vskip .2cm
\noindent
The algebra $\A_\alpha$ of entire analytic elements is norm dense in $\A$.}

States are, by definition, positive, linear and normalized functionals over $\A$. 
It is an advantage of the abstract setting that the thermal equilibrium
states can be distinguished among the set of all (physical) states 
by first principles such as time invariance,
stability against small perturbations or passivity properties (see Ref.\ 6 and 7).
Adding a few technical assumptions one usually ends up (see Ref.\ 8 and 9) with a precise 
mathematical selection criterion: the KMS condition. Today
the KMS condition is generally accepted as {\sl the} appropriate mathematical
criterion for equilibrium. But only recently, Buchholz and Junglas have shown that the 
characterization of equilibrium states by the KMS condition applies even
to a large class of relativistic models.$^{10}$ 

{\it Lorentz invariance} is always broken by a KMS state.$^{11,12}$  
A KMS state might also break {\it spatial translation} or 
{\it rotation invariance}, but
the maximal propagation velocity of signals, which is characteristic for a relativistic theory,
will not be affected by such
a lack of symmetry. It was first recognized by Bros and Buchholz 
that a finite maximal propagation velocity
of signals basically implies that the KMS states of a relativistic QFT have stronger 
analyticity properties in configuration space than those imposed by the traditional 
KMS condition.$^2$ These properties are summarized in the following

\DefII{A state $\omega_\beta $ satisfies the {\it relativistic KMS condition}   
at inverse temperature $\beta > 0$ if and only
if there exists some positive timelike vector $e \in V_+$, $e^2 = 1$, such that 
for every pair of elements $a, b$ of $\A$ there exists a function~$F_{a,b}$ which is
analytic in the domain
\# { -{\cal T}_{\beta e /2} \times {\cal T}_{\beta e /2},}
where ${\cal T}_{\beta e /2} = \{ z \in \C : \Im z \in V_+ \cap ( \beta e / 2 + V_- ) \}$ is a
tube, and  continuous at the boundary sets 
$\R^4 \times \R^4$ and $\bigl(\R^4 - {i \over 2} \beta e \bigr) \times \bigl(\R^4 
+ {i \over 2} \beta e \bigr)$ with boundary values given by
\& 
{ F_{a,b} (x_1, x_2) & = \omega_\beta \bigl( \alpha_{x_1} ( a) \alpha_{x_2} (b) \bigr)  
\cr
F_{a,b} \Bigl( x_1 - {i \over 2} \beta e, 
x_2 + {i \over 2} \beta e \Bigr) & = \omega_\beta \bigl( \alpha_{x_2} (b)
\alpha_{x_1} ( a) \bigr)   \qquad \forall x_1, x_2 \in \R^4 .}  
}

The relativistic KMS condition can be understood as a remnant of the relativistic 
spectrum condition in the vacuum sector. It has been rigorously established$^2$
for the KMS states constructed by Buchholz and Junglas.$^{10}$ 
In this letter we will show that together with the condition of additivity (see (13) below)
it implies that the KMS state has the Reeh-Schlieder property.

Once a relativistic 
KMS state $\omega_\beta$ for some inverse temperatur $\beta$ is fixed, 
the well known GNS-construction provides a Hilbert space~$\H_\beta$, a cyclic 
vector $\Omega_\beta \in \H_\beta$ and
a `thermal representation'~$\pi_\beta$ of~$\A$ 
such that 
\# { \omega_\beta (a) = \bigl( \Omega_\beta \, , \, \pi_\beta (a) \Omega_\beta \bigr)  \qquad 
\forall a \in \A . }
Due to the KMS condition the vector $\Omega_\beta$ is not 
only cyclic for ${\cal R}_\beta :=
\pi_\beta (\A)''$ but also separating. (Note that a priori the relativistic KMS 
condition only applies to elements of $\A$ and 
in general it will not extend to ${\cal R}_\beta$.) Thus any state, which is normal  
w.r.t.\ $\pi_\beta$, is a vector state (see Ref.\ 9, 2.5.31).

\Rem{Our main concern may be formulated as follows: Given a state $\omega \colon \A \to \C$,
can we find an element $a_\omega$ in $\A (\O)$ (representing a strictly local operation in $\O$)
such that $\| \omega -  \hat{\omega} \| < \epsilon$,
where $\hat{\omega} $ is identified with the normal state induced by 
$\Omega_{\hat{\omega}} := \pi_\beta (a_\omega) \Omega_\beta$? Obviously, it is sufficent to 
prove that $\Omega_{\hat{\omega}}$ can be chosen arbitrarily  
close to $\Omega_{\omega}$ in the Hilbert space topology, where $\Omega_{\omega}$ is 
the state vector associated with the state $\omega$. This will be a direct consequence of
Theorem III.9.}

The representation $\pi_\beta$ assigns to any $\O \subset \R^4$ a von Neumann algebra
\# 
{ \qquad {\cal R}_\beta (\O) = \pi_\beta \bigl( \A(\O) \bigr)''. }
The weak closure respects the local structure, i.e.,
\# { {\cal R}_\beta (\O_1 ) \subset  {\cal R}_\beta ( \O_2)' \quad \hbox{\rm for} \quad \O_1 
\subset \O_2' .}
Note that ${\cal R}_\beta ( \O)'$ denotes the commutant of 
${\cal R}_\beta ( \O)$ in the algebra $\B(\H_\beta)$ of all
bounded operators on $\H_\beta$. We emphasize that ${\cal R}_\beta ( \O)'$ includes both 
the algebra
${\cal R}_\beta'$, which itself is isomorphic to ${\cal R}_\beta$,
and ${\cal R}_\beta (\O')$ as subalgebras. 

\DefII{The net $\O \to {\cal R}_\beta (\O)$ is called {\it additive},  
if
\# { \cup_{i \in I} \O_i = \O \Rightarrow \vee_{i \in I} {\cal R}_\beta (\O_i) 
= {\cal R}_\beta (\O)  .}
Here $I$ is some index set and $\vee_i {\cal R}_\beta (\O_i)$ denotes the von Neumann  
algebra generated by the algebras ${\cal R}_\beta (\O_i)$, $i \in I$.}

If $\omega_\beta$ is locally normal w.r.t.\ the vacuum representation, then
additivity in the vacuum sector and additivity in the thermal sector are
equivalent. As is well known, additivity in the vacuum sector can be proven, 
if the net of local algebras is constructed from a Wightman field theory.

\vskip 1cm

\Hl{The Reeh-Schlieder Property}

\noindent
We start with the following 
adapted and simplified version of Glaser's Theorem 1 
(see Ref.\ 13, see also Ref.\ 14, 15):

\Th{(Glaser): Let $a \in \A$ and let $F_{a^*,a}$ denote the function introduced in (9). 
The following properties are equivalent:
\vskip .3cm
\halign{ \indent #  \hfil & \vtop { \parindent =0pt \hsize=34.8em
                            \strut # \strut} \cr 
i.)     & There exists an open neighborhood ${\cal V}$ of $0$ in $\R^4$ 
and a point $z_1 \in {\cal T}_{\beta e /2}$ such that 
$z_1 + {\cal V} \subset  {\cal T}_{\beta e /2}$ and such that 
for each complex-valued testfunction $f$ with support in ${\cal V}$
\# { \int_{\r^4 \times \r^4} {\rm d}^4 y_1 {\rm d}^4 y_2 \, \, 
F_{a^*,a} (y_1 + \bar z_1, y_2 + z_1) \overline{ f (y_1) } f(y_2)  \ge 0.}    
\cr
ii.)     & There exists a sequence 
$\bigl\{ f_a^{(n)} \colon {\cal T}_{\beta e /2} \to \C  \bigr\}_{n \in \n} $ of 
functions holomorphic in ${\cal T}_{\beta e /2}$ such that
for $(z_1, z_2) \in -{\cal T} \times {\cal T}$
\# { F_{a^*,a} (z_1, z_2) 
= \sum_{n \in \n} \overline {  f_a^{(n)} ( \bar z_1)} f_a^{(n)} (z_2)}    
holds in the sense of uniform convergence on every compact subset of 
$-{\cal T}_{\beta e /2} \times {\cal T}_{\beta e /2}$. \cr}
}

The next step is to show that condition i.)\  is indeed satisfied, if $\omega_\beta$ is a 
relativistic KMS state:

\Prop{Let $\omega_\beta$ be a state which satisfies the {\it relativistic KMS condition}   
at inverse temperature $\beta > 0$ and let ${\cal V}$ be an open neighborhood  
of $0$ in $\R^4$. 
\vskip 0cm
It follows that for each complex-valued test function $f$ with support in ${\cal V}$
\# { \int_{\r^4 \times \r^4} {\rm d}^4 y_1 {\rm d}^4 y_2 \, \, 
F_{a^*,a} (y_1 - i \kappa e, y_2 + i  \kappa e) \overline{ f (y_1)} \, f(y_2)  \ge 0 }    
for all $0 <  \kappa < \beta /2$. Here $e$ denotes the unit vector in the time direction
distinguished by the relativistic KMS condition.}

\Pr{Let $a \in \A_\alpha$ be an entire analytic element for the translations. Put 
\# { \Psi_f := \int_{\cal V} {\rm d}^4 y_1 \, \, 
f(y_1) \alpha_{y_1} \bigl(\alpha_{i  \kappa e} (a) \bigr) 
\Omega_\beta \in \H_\beta. }
Exploring the definiton (9) of $F_{a^*,a}$ one finds
\# { \int_{\r^4 \times \r^4} {\rm d}^4 y_1 {\rm d}^4 y_2 \, \, 
F_{a^*,a} (y_1 - i \kappa e,  y_2 + i  \kappa e) \overline { f (y_1) } f(y_2)  
= \| \Psi_f \|^2 \ge 0.}    
For general $a \in \A$, choose a sequence $\{ a_n \in \A_\alpha \}_{n \in \n}$  such that
\# { \| a_n \| \le \| a \| \qquad \hbox {and} \qquad \pi_\beta (a_n) \Omega_\beta \to 
\pi_\beta (a) \Omega_\beta \quad \hbox {as} \quad n \to \infty.}
Now define, for $y_1, y_2 \in \R^4$ and $0 <  \kappa < \beta /2$,
\# {F_n (y_1 - i \kappa e,  y_2 + i  \kappa e) 
:= F_{a_n^*,a_n} (y_1 - i \kappa e,  y_2 + i  \kappa e) .}
The Three-line Theorem (see Ref.\ 9, 5.3.5) implies that 
\# { \Bigl| F_n (y_1 - i \kappa e,  y_2 + i  \kappa e) -  
F_m (y_1 - i \kappa e,  y_2 + i  \kappa e) \Bigr| }
assumes its maximum value on the boundary of its domain and for $\kappa = 0, \beta /2 $, the
boundary values, the relativistic KMS condition yields
\& {\Bigl| F_n (y_1 - i \kappa e, &  y_2 + i  \kappa e) -  
F_m (y_1 - i \kappa e ,  y_2 + i  \kappa e ) \Bigr| 
\cr
& \le 
\max \Bigr\{ \sup_{y_1, y_2 \in \r^4} 
\bigl| \omega_\beta \bigl(\alpha_{y_1} (a^*_n) \alpha_{y_2} (a_n) \bigr) - 
\omega_\beta \bigl(\alpha_{y_1} (a^*_m) \alpha_{y_2} (a_m) \bigr) \bigr|,
\cr
&
\qquad
\qquad
\qquad
\sup_{y_1, y_2 \in \r^4} \bigl| \omega_\beta \bigl(\alpha_{y_2} (a_n) 
\alpha_{y_1} (a_n^*) \bigr) - 
\omega_\beta \bigl(\alpha_{y_2} (a_m) \alpha_{y_1} (a_m^*) \bigr) \bigr| \Bigr\}
\cr
& \le \sup_{y_1, y_2 \in \r^4} 
\bigl| \omega_\beta \bigl(\alpha_{y_1} (a^*_n) \alpha_{y_2} (a_n) \bigr) -
\omega_\beta \bigl(\alpha_{y_1} (a^*_n) \alpha_{y_2} (a_m) \bigr) \bigr|
\cr
&
\quad +  \sup_{y_1, y_2 \in \r^4} 
\bigl| \omega_\beta \bigl(\alpha_{y_1} (a^*_n) \alpha_{y_2} (a_m) \bigr) -
\omega_\beta \bigl(\alpha_{y_1} (a^*_m) \alpha_{y_2} (a_m) \bigr) \bigr| 
\cr
& \quad
+ \sup_{y_1, y_2 \in \r^4} \bigl| \omega_\beta \bigl(\alpha_{y_2} (a_n) 
\alpha_{y_1} (a_n^*) \bigr) - \omega_\beta \bigl(\alpha_{y_2} (a_n) 
\alpha_{y_1} (a_m^*) \bigr) \bigr|
\cr
& \quad
+ \sup_{y_1, y_2 \in \r^4} \bigl| \omega_\beta \bigl(\alpha_{y_2} (a_n) 
\alpha_{y_1} (a_m^*) \bigr)  -
\omega_\beta \bigl(\alpha_{y_2} (a_m) \alpha_{y_1} (a_m^*) \bigr) \bigr| \Bigr\}
\cr
& \le  2 \, \| a \| \sup_{y \in \r^4} 
\bigl\| \pi_\beta \bigl(\alpha_{y} (a_n -a_m) \bigr) 
\bigr\|  
+ 2 \, \| a \| \sup_{y \in \r^4} 
\bigl\| \pi_\beta \bigl(\alpha_{y} (a_m^* -a_n^*)\bigr) 
\bigr\|  .}
In the last inequality we have used $ \| a_n \| = \| a_n^* \| \le \| a \| $ 
and $\| \Omega_\beta \|=1$.
Strong continuity of $\alpha$ now implies that 
$\{ F_n \}_{n \in \n}$ is a Cauchy sequence uniformly on $\overline {\cal U}$, where
\# { {\cal U} :=  \{ (y_1 - i \kappa e ,  y_2 + i  \kappa e )  : y_1, y_2 \in \R^4,
0 <  \kappa < \beta /2  \}   .}
The limit function $F_\infty$ is therefore continuous and bounded 
on $\overline {\cal U}$ and analytic
in ${\cal U}$. Moreover, 
\# { F_\infty (y_1, y_2) = F_{a^*,a} (y_1,  y_2 ) \qquad \hbox{for} \quad y_1, y_2 \in \R^4.}
Thus, due to their analyticity properties, the functions $F_\infty$ and $F_{a^*,a}$ must
coincide on $ {\cal U}$. It follows that
\& { \int_{\r^4 \times \r^4} {\rm d}^4 y_1 & {\rm d}^4 y_2 \, \, 
F_{a^*,a} (y_1 - i \kappa e , y_2 + i  \kappa e ) \overline{ f (y_1) } f(y_2)  =
\cr
& 
\lim_{n \to \infty}  \int_{\r^4 \times \r^4} {\rm d}^4 y_1 {\rm d}^4 y_2 \, \, 
F_n (y_1 - i \kappa e ,  y_2 + i  \kappa e ) \overline { f (y_1) }f(y_2)
\ge 0.}    
}

The crucial step in the proof of the Reeh-Schlieder property 
is now summarized in the following

\Prop{For each $a \in \A$ the vector valued function $\Phi_a \colon \R^4 \to \H_\beta$,
\# { x \mapsto \pi_\beta \bigl( \alpha_{x} (a) \bigr) \Omega_\beta  }  
can be analytically continued from the real axis into the domain
${\cal T}_{\beta e / 2 }$
such that it is weakly continuous for $\Im z \searrow 0$. }

\Pr{Let $a, b \in \A$ with $ \| a \|=1$. Because of
\# { \overline{ \pi_\beta (\A) \Omega_\beta} = \H_\beta}
the set of vectors ${\cal S} := \{ \pi_\beta (b) \Omega_\beta : b \in \A \}$ is dense 
in $\H_\beta$. Moreover, according to Theorem III.1.ii)\ there exists a sequence 
$\bigl\{ f_a^{(n) }\colon {\cal T}_{\beta e /2} \to \C   \bigr\}_{n \in \n} $ of 
functions holomorphic in ${\cal T}_{\beta e /2}$ which satisfies (15). This allows us to
consider --- for $z \in {\cal T}_{\beta e / 2 }$ and $a \in \A$ fixed --- the map
$\hat \phi_{a,z} \colon {\cal S} \to \C$ 
\# { \pi_\beta (b) \Omega_\beta  \mapsto 
\sum_{n \in \n} \overline {  f_a^{(n)} ( \bar z)} \, f_b^{(n)} (0) .}    
(Recall that $\Omega_\beta$ is separating for $\pi_\beta (\A)$. 
Hence the map $b \mapsto \pi_\beta (b) \Omega_\beta $ is injective and consequently
the map $\pi_\beta (b) \Omega_\beta \mapsto f_b^{(n)} (0)$ is well-defined.)
Using $\|a \| \le 1$,
\# { \sum_{n \in \n} \bigl| f_a^{(n)} (z) \bigr|^2  = F_{a^*,a} (\bar z, z) ,}    
and the Schwarz inequality, we find 
\# { \Bigl| \sum_{n \in \n} \overline {  f_a^{(n)} ( \bar z)} \, f_b^{(n)} (0) \Bigr|^2 
\le F_{a^*,a} (\bar z, z) \cdot \| \pi_\beta (b) \Omega_\beta \|^2.} 
By the Hahn-Banach Theorem the map $\hat \phi_{a,z} \colon {\cal S} \to \C$ extends to 
a (bounded) continuous linear functional
$\phi_{a,z}$ on $\H_\beta$. The Riesz Lemma ensures that
there exists a vector $\Phi_a (z) \in \H_\beta$ 
such that
\# { \phi_{a,z} (\Psi) = \bigl(\Phi_a (z)  \, , \, \Psi\bigr) \qquad \forall \Psi \in \H_\beta.}
The map
\# { z \mapsto \Phi_a (z) }
is analytic for $z \in {\cal T}_{\beta e / 2 }$. (This can be shown by an approximation
argument similar to the one given in the proof of Proposition III.2.)
As can be seen more easily, the map (32) is also weakly continuous at the boundary set $\Im z = 0$,
where it satisfies
\# { \Phi_a (x) = \pi_\beta \bigl( \alpha_{x} (a) \bigr) \Omega_\beta  
\qquad \forall x \in \R^4.}  
}

Although we will not directly use it, we believe that it is worthwhile to spell out the following

\Cor{Let $a, b \in \A$ and let $\Phi_a $, $\Phi_b$ denote the associated vector valued functions
introduced in (32).
It follows that 
\# { F_{a^*,b} (\bar z_1,z_2) = \bigl( \Phi_a (z_1)  \, , \, \Phi_b (z_2) \bigr) 
}
for all $z_1, z_2 \in {\cal T}_{\beta e / 2 }$. Here $F_{a^*,b}$ denotes the 
analytic function introduced in (9).}

\Pr{The l.h.s.\ as well as the r.h.s.\ defines a holomorphic function on 
\# {  -{\cal T}_{\beta e /2} \times {\cal T}_{\beta e /2} \subset \C^4 \times \C^4 .}
Moreover, for $\Im z_1 \searrow 0$ and $\Im z_2 \searrow 0$ we find
\& { \bigl( \Phi_a (x_1)  \, , \, \Phi_b (x_2) \bigr) & = 
\bigl( \pi (\alpha_{x_1} (a)) \Omega_\beta   \, , \, \pi (\alpha_{x_2} (b)) \Omega_\beta \bigr)
\cr
&
= \omega_\beta  \bigl( \alpha_{x_1} (a^*) \alpha_{x_2} (b)\bigr) = F_{a^*,b} (x_1,x_2) }
for all $x_1, x_2 \in \R^4$. Applying the Edge-of-the-Wedge Theorem we conclude that the
l.h.s.\ and the r.h.s.\ in (34) describe the same analytic function.}

What remains to be proven in order to establish the Reeh-Schlieder property
is fairly standard. Borchers and Buchholz$^{16}$ recently 
gave a nice and transparent formulation of this final part of the argument and 
therefore we will simply reproduce their formulation here, up to minor notational differences.

\DefIII{Let $\O$ be any open region. The $*$-algebra  $\B(\O)$ is defined as the set of operators 
$b \in \A(\O)$ for which there exists some neighborhood ${\cal N} \subset \R^4$ of the origin
such that 
\# { \alpha_x (b) \in \A(\O)  \qquad \forall x \in {\cal N}, }
where the neighborhood ${\cal N}$ may depend on $b$.}

$\B(\O)$ is a $*$-algebra and 
\# { \A(\O_\circ) \subset \B(\O) }
for any region $\O_\circ$ whose closure lies in the interior of~$\O$.

\Lm{Let $\Psi \in \H_\beta$ be a vector with the property that
\# { \bigl( \Psi \, , \, \pi_\beta (b) \Omega_\beta \bigr) = 0 \qquad \forall b \in \B(\O). }
It follows that for each $b \in \B(\O)$ the function 
\# { \R^4 \ni x \mapsto \bigl( \Psi \, , \, \pi_\beta \bigl(\alpha_x(b) \bigr) \Omega_\beta \bigr) }
vanishes.}

\Pr{Let $b \in \B(\O)$ and let ${\cal N}$ as in (37). It follows from the definition of $\B(\O)$
and the geometrical action (5) of the translations that there exists some 
$\epsilon >0$, which may depend on~$b$, such that
\# { \alpha_x (b) \in \B(\O) \quad  \hbox{for} \quad  |x| < \epsilon.}
On the other hand the function
\# { \R^4 \ni x \mapsto \pi_\beta \bigl( \alpha_x (b) \bigr) \Omega_\beta }
extends analytically to some vector-valued function in the domain
${\cal T}_{\beta e / 2 }$ by Pro\-po\-si\-tion~III.3.
Combining (42) with (41) and (39) we find that 
\# { \bigl( \Psi \, , \, \pi_\beta \bigl( \alpha_x (b) \bigr) \Omega_\beta \bigr) = 0 } 
for all $x \in \R^4$. }

\Lm{Assume that the additivity assumption (13) holds. It follows that
\# {  \vee_{x \in \r^4} \pi_\beta \Bigl(\alpha_x \bigl( \B(\O) \bigr) \Bigr)''   
= {\cal R}_\beta  .}
Once again, $\vee_{x \in \r^4} \pi_\beta \bigl(\alpha_x ( \B(\O) ) \bigr)''$
denotes the von Neumann algebra generated by the algebras
$\pi_\beta \bigl(\alpha_x ( \B(\O)  \bigr)''$, $x \in \R^4$.}

\Pr{Let $\O_\circ$ be an open subset of $\O$ such that its closure 
$\overline {\O_\circ}$ is contained in the interior
of~$\O$. It follows that
\# {  \vee_{x \in \r^4} \pi_\beta \Bigl(\alpha_x \bigl( \B(\O) \bigr) \Bigr)'' \supset
\vee_{x \in \r^4} \pi_\beta \Bigl(\alpha_x \bigl( \A(\O_\circ) \bigr) \Bigr)''  .}
Combining (5) with (13) we conclude that the r.h.s.\ equals ${\cal R}_\beta$.} 

\Cor{Let $\Psi \in \H_\beta$ be a vector with the property that
\# { \bigl( \Psi \, , \, \pi_\beta (b) \Omega_\beta \bigr) = 0 \qquad \forall b \in \B(\O). }
It follows that $\Psi = 0$.}

\Pr{First, we apply Lemma III.6 and conclude from (40) that 
\# { \Psi \bot  \vee_{x \in \r^4} 
\pi_\beta \bigl(\alpha_x \bigl( \B(\O) \bigr) \bigr) \Omega_\beta.}
Then we recall that the orthogonal complement of 
$\pi_\beta \bigl(\alpha_x \bigl( \B(\O) \bigr) \bigr) \Omega_\beta$ is closed,
therefore it coincides with the orthogonal complement of 
$\pi_\beta \bigl(\alpha_x \bigl( \B(\O) \bigr) \bigr)'' \Omega_\beta$.
Hence Lemma III.7 implies
\# { \Psi \bot  \overline {  {\cal R}_\beta \Omega_\beta  } . }
By construction $\overline {  {\cal R}_\beta \Omega_\beta }= \H_\beta$,
thus $\Psi = 0$.}

We will now show that for every vector $\Phi \in \H_\beta$ 
there exists an operator in $\pi_\beta \bigl( \A (\O)\bigr)$, which, when applied to
$\Omega_\beta$, generates a vector which is arbitrarily close to~$\Phi$:  

\Th{Consider a QFT as specified in Section 2 and let $\omega_\beta$ 
be a state, which satisfies the 
relativistic KMS condition. If the additivity assumption (13) holds, then
\# { \H_\beta = \overline { \pi_\beta \bigl( \A (\O) \bigr) \Omega_\beta},    }
for any open space--time region $ \O  \subset \R^4$. Moreover,
if the spacelike complement of $\O$ is not empty, then
$\Omega_\beta$ is separating for $ {\cal R}_\beta (\O )$.  } 

\Pr{We have to show that the orthogonal complement of
$\overline { \pi_\beta \bigl( \A (\O) \bigr) \Omega_\beta}$ vanishes.
Assume that 
\# { \Psi \bot  \pi_\beta \bigl( \A (\O) \bigr) \Omega_\beta  .}
Obviously, this implies  
\# { \Psi \bot  \pi_\beta \bigl( \B (\O) \bigr) \Omega_\beta  }
and then Corollary III.8 yields $\Psi = 0$.  
On the other hand, if the spacelike complement $\O'$ of $\O$ is not empty, then 
$\Omega_\beta$ is cyclic for~${\cal R}_\beta (\O') \supset \pi_\beta \bigl( \A(\O') \bigr)$.
Since ${\cal R}_\beta (\O)' \supset {\cal R}_\beta (\O') $, this implies that 
$\Omega_\beta$ is cyclic for ${\cal R}_\beta (\O)'$ and therefore
separating for~${\cal R}_\beta (\O)$.}

Similar to the situation where $\beta = \infty$, in the so-called 
vacuum sector, $\Omega_\beta$  shares the Reeh-Schlieder property with 
a large class of vectors in $\H_\beta$. 

\Th{There exists a dense set ${\cal D}_\alpha \subset \H_\beta$, 
such that for all 
$\Psi \in {\cal D}_\alpha$  
\# { \H_\beta = \overline { \pi_\beta \bigl( \A (\O) \bigr) \Psi},    }
where $ \O  \subset \R^4$ is again an arbitrary open space--time region.}

\Pr{A set ${\cal D}_\alpha \subset {\cal R}_\beta \Omega_\beta$ 
of suitable entire analytic vectors in $\H_\beta$ may be specified by putting
\# {  {\cal D}_\alpha = \Bigl\{ \Bigl( \1 - { \pi_\beta(a) \over 2 \, \| a \| } \Bigr) 
\Omega_\beta :  a \in \A_\alpha \Bigr\} .}
Note that ${\cal D}_\alpha$ is dense in $\H_\beta$:
\# { \overline {{\cal D}_\alpha} = \overline {\pi_\beta (\A_\alpha) \Omega_\beta}=
\overline {\pi_\beta (\A) \Omega_\beta} = \H_\beta.}
The essential step is to show that for 
arbitrary $b \in \A$ the function
\# { \R^4 \ni x \mapsto \pi_\beta \bigl( \alpha_x (b) \bigr) \Psi }
extends to some analytic vector-valued function in the domain
${\cal T}_{\beta e / 2 }$. The reader is invited to check that Theorem III.1 and Proposition 
III.2 can easily be adapted and that the proofs given  
remain valid if we replace~$\Omega_\beta$ by some vector 
$\Psi \in {\cal D}_\alpha$. Finally, we note that
$\bigl( \1 - a  /  2 \, \| a \|   \bigr)$
is invertible in $\A$, thus
\# { \pi_\beta (\A) \Bigl( \1 - { \pi_\beta(a) \over 2 \, \| a \| } \Bigr) 
\Omega_\beta = \pi_\beta (\A) \Omega_\beta.}
We conclude that $\overline { \pi_\beta (\A) \Psi } = \H_\beta$, which ensures that  
the arguments given in the proof of Corollary III.8 apply also in this slightly more general case.}

\vskip 1cm

\Hl{Outlook}

\noindent
Although both quantum statistical mechanics as well as quantum field theory 
can nicely be formulated in terms of operator algebras, little 
 Of course, one should
not forget to mention the beautiful progress that has recently is known about the thermal states of 
a relativistic system. 
been achieved in the Wightman approach to
thermal field theory (see, e.g., Ref.\ 17).  
In fact, only recently the relativistic KMS condition, which provides the necessary substitute 
for the spectrum condition, was formulated. As we have demonstrated, it 
allows us to treat the thermal theory independently from the vacuum theory. 
In a series of forthcoming papers by the author
basic results like the Cluster Theorem, the Schlieder property and the Borchers property
have been derived. The nuclearity condition, which distinguishes theories with decent phase-space
properties, was used to derive the split property, 
which expresses a strong form of statistical independence of spacelike separated measurements.
A rather involved argument, based on a rigorous version of what is commonly called `doubling the
degrees of freedom' in thermal field theory, establishes the `convergence of local charges' 
in the thermal sector. These results use the Reeh-Schlieder property as a crucial input; and
without the present result one would have to recourse to the physically sound but unproven
assumption that relativistic KMS states are locally normal w.r.t.\ the vacuum representation.

Many other bricks are still missing in the wall, for instance a thermal 
Jost-Lehmann-Dyson representation. Scattering theory in the 
thermal context is one of the major challenges; in fact, the author would like to emphasize
that the severe problems encountered in perturbation theory open up a fair chance for
the operator algebraic approach to attract some interest from outside, 
provided it can offer some progress on this topic in time.

\vskip .5 cm

\noindent
{\it  Acknowledgements.\/}
\noindent
This work was supported by the CEE. The author wants to thank D.~Buchholz for  
pointing out the relevance of Glaser's Theorem and the referee for numerous improvements. 
Kind hospitality of the Dipartimento di 
Matematica, Universita di Roma ``Tor Vergata" and the Erwin Schr\"odinger Institute (ESI), 
Vienna, is gratefully acknowleged.

\vskip 1cm

\noindent
{\fourteenrm References}
\nobreak
\vskip .3cm
\nobreak
\halign{   &  \vtop { \parindent=0pt \hsize=33em
                            \strut  # \strut} \cr 
\BOOK
{1}
{Junglas, P.}   {Thermodynamisches Gleichgewicht und Energiespektrum in der 
                 Quantenfeldtheorie} 
                {Dissertation, Hamburg}
                {1987}
\REF
{2}
{Bros, J.\ and Buchholz, D.}      {Towards a relativistic KMS condition}
                                  {Nucl.\ Phys.\ B} 
                                  {429} {291--318}
                                  {1994}
\REF
{3}
{J\"akel, C.D.}                   {Decay of spatial correlations in thermal states}
                                  {Ann.\ l'Inst.\ H.\ Poincar\'e} 
                                                      {69} {425--440}
						                                                {1998}
\BOOK
{4}
{Haag, R.}    {Local Quantum Physics: Fields, Particles, Algebras} 
              {Springer-Verlag, Berlin-Heidelberg-New York} 
              {1992}
\BOOK
{5}
{Sakai, S.}     {Operator Algebras in Dynamical Systems} 
                {Cambridge University Press, Cambridge-New York-Port Chester-Melbourne-Sydney}  
                {1991}
\REF
{6}
{Haag, R., Kastler, D.\ and Trych-Pohlmeyer, E.B.}    {Stability and equilibrium states}
                                                      {\CMP} 
                                                      {38} {173--193}
						                                                {1974}
\REF
{7}
{Pusz, W., and Woronowicz, S.L.}   {Passive states and KMS states for general 
               quantum systems}
              {\CMP}
              {58}    {273--290}
                          {1978}
\REF
{8}
{Haag, R., Hugenholtz, N.M.\ and Winnink, M.}
                          {On the equilibrium states in quantum statistical mechanics}  
                          {\CMP}	
                          {5}	{215--236}
                          {1967}
\BOOK
{9}  
{Bratteli, O.\ and Robinson, D.W.} {Operator Algebras and Quantum Statistical Mechanics~I, II} 
                                  {Sprin\-ger-Verlag, New York-Heidelberg-Berlin} 
                                  {1981}
\REF
{10}
{Buchholz, D.\ and Junglas, P.}   {On the existence of equilibrium states in local 
                                   quantum field theory} 
                                  {\CMP} 
                                  {121} {255--270}
                                  {1989}
\REF
{11}
{Narnhofer, H.}   {Kommutative Automorphismen und Gleichgewichtszust\"ande}
                  {Act.\ Phys.\ Austriaca}
                  {47}   {1--29} 
                  {1977}
\REF
{12}
{Ojima, I.}   {Lorentz Invariance vs. Temperature in QFT}
              {\LMP}
              {11}    {73--80}
              {1986}
\BOOK
{13}
{Glaser, V.}      {The positivity condition in momentum space}
                       {In {\sevensl Problems in Theoretical Physics.
                        Essays dedicated to N.N.\ Bogoliubov.} 
                        D.I. Bolkhintsev et al.\ eds.\ Moscow, Nauka}
                       {1969}
\REF
{14}
{Glaser, V.}      {On the equivalence of the Euclidean and Wightman formulation of 
                   field theory}
                          {\CMP}	
                          {37}	{257--272}
                          {1974}
\REF
{15}
{Bros, J., Epstein, H.\ and Moschella, U.}      {Analyticity properties and thermal 
                                   effects for general quantum field theory on de Sitter 
                                   space--time}
                                  {\CMP} 
                                  {196} {535--570}
                                  {1998}
\REF
{16}
{Borchers, H.J.\ and Buchholz, D.}       {Global properties of vacuum states in de Sitter space}
                                  {Ann.\ l'Inst.\ H.\ Poincar\'e} 
                                                      {70} {23--40}
						                                                {1999}
\REF
{17}
{Bros, J.\ and Buchholz, D.}      {Axiomatic analyticity properties and representations
                                   of particles in thermal quantum field theory}
                                  {Ann.\ Inst.\ H.\ Ponicar\'e} 
                                  {64} {495--521}
                                  {1996}
\cr}

\bye